\documentclass[12pt,dvipsnames]{article}

\usepackage{iftex}
\ifLuaTeX
\usepackage{polyglossia} 
\usepackage{fontspec} 
\setmainlanguage{english}
\usepackage[nosort]{cite}
\else 
\usepackage[utf8]{inputenc}
\usepackage[T1]{fontenc}
\usepackage[nosort]{cite}
\usepackage[english]{babel} 
\fi

\usepackage[top=2cm,bottom=2cm,outer=2.5cm,inner=2.5cm,marginparwidth=2.5cm]{geometry}
\usepackage{fvextra} 
\usepackage{authblk} 

\ifLuaTeX
\usepackage[unicode]{hyperref}
\usepackage{pdftexcmds} 
\makeatletter
\let\pdfstrcmp\pdf@strcmp
\makeatother
\else
\usepackage{hyperref}
\fi

\usepackage{amsmath}
\usepackage{amsfonts}
\usepackage{amssymb}
\usepackage{tensor} 
\usepackage{braket} 
\usepackage{empheq} 
\usepackage{romanbar} 
\usepackage{dsfont} 
\usepackage{stmaryrd} 

\usepackage{lmodern} 
\usepackage{caption}
\usepackage{subcaption} 
\usepackage{titlesec} 
\usepackage{fancyhdr} 
\usepackage{abstract} 
\usepackage[super]{nth}
\usepackage[colorinlistoftodos,prependcaption,textsize=tiny]{todonotes} 
\usepackage{tocloft} 

\usepackage{graphicx} 
\usepackage{multirow} 
\usepackage{array} 
\usepackage{tikz}
\usepackage[compat=1.1.0]{tikz-feynman} 
\usepackage{pgf}
\usepackage{pgfplots} 
\usepackage{dynkin-diagrams}
\usepackage{hhline}

\usetikzlibrary{arrows,shapes,calc,intersections,fadings,positioning,shadows.blur,decorations.pathreplacing,external,backgrounds}
\usepackage{currfile} 

\immediate\write18{mkdir -p TikZ}     
\tikzexternaldisable
\newcommand*\setmyname{
\expandafter\tikzsetfigurename\expandafter{\currfilebase-}%
}
\AtBeginEnvironment{tikzpicture}{\setmyname}

\hypersetup{
pdfstartview={FitH},
pdftitle={BPS minimal volumes},
pdfauthor={N. Drukker, M. Trépanier},
pdfsubject={},
pdfcreator={},
pdfkeywords={},
colorlinks=true,
citecolor=MidnightBlue,
linkcolor=MidnightBlue,
urlcolor=MidnightBlue
}

\numberwithin{equation}{section}
\newcommand\frontmatter{%
\clearpage
\pagenumbering{roman}
}

\newcommand\mainmatter{%
\clearpage
\pagenumbering{arabic}
}

\usepackage{mathtools}

\let\Re\undefined
\DeclareMathOperator{\Re}{Re}
\let\Im\undefined
\DeclareMathOperator{\Im}{Im}

\newcommand{\vev}[1]{\left\langle #1 \right\rangle}

\DeclareMathOperator{\vol}{vol}

\newcommand{\diff}{\mathrm{d}}

\def\bC {\mathbb{C}}

\def\bH {\mathbb{H}}

\def\bL {\mathbb{L}}

\def\bR {\mathbb{R}}

\def\cN{{\mathcal{N}}}

\def\cO{{\mathcal{O}}}

\newcommand{\bea}{\begin{eqnarray}}
\newcommand{\eea}{\end{eqnarray}}
\newcommand{\beq}{\begin{equation}}
\newcommand{\eeq}{\end{equation}}
\newcommand{\bal}{\begin{equation}\begin{aligned}}
\newcommand{\eal}{\end{aligned} \end{equation}}

\title{BPS surface operators and calibrations}
\author[1]{Nadav Drukker\thanks{\href{mailto:nadav.drukker@gmail.com}{nadav.drukker@gmail.com}}}
\author[1,2]{Maxime
Tr\'epanier\thanks{\href{mailto:trepanier.maxime@gmail.com}{trepanier.maxime@gmail.com}}}
\affil[1]{\it Department of Mathematics, King's College London,\protect\\London, WC2R
2LS, United Kingdom}
\affil[2]{\it Perimeter Institute for Theoretical Physics,\protect\\
Waterloo, ON N2L 2Y5, Canada}
\date{}

\begin{document}

\frontmatter
\maketitle

\begin{abstract}
We present here a careful study of the holographic duals of BPS surface operators in 
the 6d ${\cal N}=(2,0)$ theory. Several different classes of surface operators have 
been recently identified and each class has a specific calibration form---a 
3-form in $AdS_7\times S^4$ whose pullback to the M2-brane world-volume is equal to the volume 
form. In all but one class, the appropriate forms are closed, so the action of the M2-brane 
is easily expressed in terms of boundary data, which is the geometry of the surface. Specifically, 
for surfaces of vanishing anomaly, 
it is proportional to the integral of the square of the extrinsic curvature. 
This can be extended to the case of surfaces with anomalies, by taking the ratio 
of two surfaces with the same anomaly.
This gives a slew of new expectation values at large~$N$ in this theory.

For one specific class of surface operators, which are Lagrangian submanifolds of 
${\mathbb R}^4\subset {\mathbb R}^6$, the structure is far richer and we find that the 
M2-branes are special Lagrangian submanifold of an appropriate six-dimensional 
almost Calabi-Yau submanifold of $AdS_7\times S^4$. This allows for an elegant treatment 
of many such examples.
\end{abstract}
\thispagestyle{empty}

\mainmatter

\section{Introduction and conclusions}
\label{sec:intro}

In a recent paper \cite{Drukker:2020bes} we presented four classes of BPS surface operators in the 
6d $\cN=(2,0)$ theory~\cite{witten:1995zh,Strominger:1995ac,ganor:1996nf, Howe:1997ue}. 
In each case the shape of the surfaces and their scalar coupling (or local 
R-symmetry breaking) have a geometric relation that guarantees that the projector 
equations for supersymmetry breaking, which have 16 independent solutions at each point, also 
share some solutions along the entire surface. These are surface 
operators in 6d analogues of BPS Wilson loops of the types 
studied in \cite{zarembo:2002an, Drukker:2007dw, Drukker:2007yx, drukker:2007qr}.

Such surfaces preserve some (conformal) Killing spinor in $\bR^6$. The holographic description 
of surface operators is in terms of an M2-brane in $AdS_7\times S^4$, which is also BPS 
and preserves the Killing spinor $\varepsilon$ related to the one on the boundary. The supergravity 
projector equation is a consequence of the $\kappa$-invariance of the M2-brane 
action~\cite{Bergshoeff:1987cm} and reads
\beq
-\frac{i}{6} \varepsilon \Gamma_{MNP} \partial_m X^M \partial_n X^N \partial_p X^P
\varepsilon^{mnp} = \varepsilon\,,
\label{eqn:kappa}
\eeq
where $X^M$ are $AdS_7\times S^4$ coordinates and 
$\varepsilon^{mnp}$ is the Levi-Civita tensor density and includes
$1/\sqrt{g}$ where $g_{mn}$ is the induced metric.

Given a preserved Killing spinor $\varepsilon$, we can construct a 
3-form~\cite{mezei:2018url}
\beq
\phi = -i \,\frac{\varepsilon \Gamma_{MNP} \varepsilon^\dagger}{\varepsilon \varepsilon^\dagger}\,
\diff X^M \wedge \diff X^N \wedge \diff X^P\,.
\label{eqn:phidef}
\eeq
By construction, the pullback of this 3-form to an M2-brane satisfying~\eqref{eqn:kappa} 
is its volume form.

If $\phi$ is closed $\diff \phi = 0$, then the form is a
calibration~\cite{harvey1982calibrated,Dymarsky:2006ve,joyce2007riemannian}; its integral is the 
same on all 3-volumes of the same homology class and is equal to the minimal
volume, so the action of the classical M2-brane.

If $\phi$ is exact then the action comes only from the boundary
and at leading order is simply a divergence proportional to the area of the surface 
operator. 
This divergent term is removed by the Legendre transform
of~\cite{drukker:1999zq,Rodgers:2018mvq,Drukker:2020dcz} or equivalently by
renormalisation~\cite{Bianchi:2001kw}. The expectation value of the surfaces is then given 
by subleading terms. If there are logarithmic divergences, the surface is anomalous and the 
coefficient of the divergence, or the anomaly is given by a universal formula unravelled in 
\cite{deser:1993yx, Berenstein:1998ij, graham:1999pm, DHoker:2008lup,DHoker:2008rje, 
Bullimore:2014upa, Gentle:2015jma, Estes:2018tnu, Jensen:2018rxu, Chalabi:2020iie,Drukker:2020swu,
Drukker:2020atp, Wang:2020xkc, Rodgers:2018mvq}. 
See \cite{Drukker:2020dcz} for a discussion of the state of the art. If there 
are no anomalies, the surface operators may have finite expectation values, which is what 
we focus on below. In cases with anomalies we can find finite ratios between expectation 
values of two different surface operators, again giving a finite computable quantity, 
see~\cite{Drukker:2022beq}.

Explicit expressions for the 3-forms for all families of BPS surface operators 
identified in~\cite{Drukker:2020bes} were presented there and are repeated below. In this paper 
we note that for three of the classes: Type-$\bR$, Type-$\bH$ and Type-$\bC$ the calibration forms 
are exact and use that to evaluate the finite part of their expectation values.

The calculation does not require to find the minimal surfaces, which is generically a hard problem, 
even in the BPS case. But it does provide explicit expressions for the classical action which we 
can compare to expressions found for all known examples. This leaves only one class of BPS surfaces, 
Type-S in the parlance of~\cite{Drukker:2020bes}, where the form is not exact and thus we cannot 
determine the expectation value directly from the boundary data without solving the equations of 
motion. This may require generalised calibrations, as 
in~\cite{Gutowski:1999iu, Gutowski:1999tu,Gutowski:1999dr}.

In the second part of this paper we look at a subclass of surfaces within Type-$\bH$ of 
Lagrangian surfaces in $\bR^4$. We find that the M2-branes are special Lagrangian surfaces 
within a 6d subspace of $AdS_7\times S^4$. We apply some of the methods of constructing 
special Lagrangian surfaces to this setting, recovering previously found solutions 
of tori, cylinder and crease~\cite{Drukker:2021vyx,Drukker:2022beq} and some 
generalisations thereof.

The rich spectrum of BPS operators of this theory~\cite{Drukker:2020bes} continue to 
provide an opening to performing explicit calculations in the $\cN=(2,0)$ theory 
in increasingly wider settings. We expect them to continue to be an ideal laboratory 
for the study of the 6d theory.

Throughout this paper we use the metric
\beq
ds^2= \frac{y}{L} \diff x_m\diff x_m
+\frac{L^2}{y^2}\diff y_I\diff y_I\,,
\label{eqn:metric}
\eeq
To avoid confusion with exponents, we use subscripts for flat indices 
$m$ and $I$ and denote $y =\sqrt{y_Iy_I}$.

\section{Surface expectation value from exact calibrations}
\label{sec:VEV}

In this section we study surfaces of Type-$\bR$, Type-$\bH$ and Type-$\bC$ that 
have finite expectation values and rely on the calibration equations to find their 
expectation values at leading order at large $N$. Surfaces of Type-$\bR$ do not have 
anomalies, while for surfaces of Type-$\bH$ and Type-$\bC$ there are constraints on their 
topology that ensure that the anomaly vanishes. This was pointed out in~\cite{Drukker:2020bes} 
and is reviewed below.

For now, let us assume that there is a surface calibrated by an exact form 
$\phi=\diff\Phi$. Then the expectation value of the surface operator is
\beq
\vev{V_\Sigma}
=\exp\left[-T_{M2}\int_V \phi\right]
=\exp\left[-\frac{N}{4\pi L^3}\int_{\Sigma=\partial V}\Phi\right],
\eeq
where the tension of the M2-brane is
\beq
T_{M2}=\frac{1}{4\pi^2\ell_P^3}=\frac{N}{4\pi L^3}\,.
\eeq
The expression for $\Phi$ depends on the type of operator, but in all cases it 
needs to be evaluated near the boundary of $AdS_7$, where we can rely on the near boundary 
solution found by Graham and Witten~\cite{graham:1999pm} relying on the 
Fefferman-Graham expansion~\cite{fefferman:2007rka}.

Using the coordinates $u$ and $v$ on the surface $\Sigma$, 
the surface is given by $x^{(0)}_\mu(u,v)$ and $n^{(0)}_I(u,v)$. For the M2-brane solution we employ 
the same coordinates and in addition $y$ from \eqref{eqn:metric}. The solution is represented 
perturbatively as
\beq
\label{large-y}
x_\mu(u,v,y)=x^{(0)}_\mu(u,v)+\frac{L^3}{y}x^{(1)}_{\mu}(u,v)+\dots\,,
\qquad
n_I(u,v,y)=n^{(0)}_{I}(u,v)+\frac{L^3}{y}n^{(1)}_I(u,v)+\dots\,.
\eeq
$x^{(1)}$ is determined \cite{graham:1999pm} to be the mean curvature vector
\beq
\label{x_1}
x_\mu^{(1)}(u,v)
= H_\mu(u,v)
=
h^{ab} \partial_a \partial_b x^{(0)}_\lambda \left( \delta_\mu^\lambda - h^{cd}
\partial_c x^{(0)}_\mu \partial_d x^{(0)}_\lambda \right)\,.
\eeq
here $h_{ab}$ is the induced metric on $\Sigma$. $n^{(1)}$ can be calculated 
following~\cite{Drukker:2020dcz}, but is not required for any of our calculations below. The 
reason is that it always appears in the combination $n^{(0)}_I n^{(1)}_I$ which vanishes, since the 
normalization $n_In_I=1$ is always assumed.

We now apply this to the various examples of BPS surfaces.

\subsection{Type-$\bR$}
\label{sec:R}

These surfaces are a product of a line in the $x_6=v$ direction 
and an arbitrary curve in the transverse direction
$x^{(0)}_I(u) \subset \bR^5$ ($I = 1, \dots, 5$). The scalar 
coupling then is chosen to be along the tangent vector to the curve
\beq
n^{(0)}_I(u,v) = \frac{\partial_u x^{(0)}_I}{|\partial_u x^{(0)}|}\,.
\label{eqn:typeR}
\eeq
Examples of surfaces in this class were previously studied 
in~\cite{Lee:2006gqa, Drukker:2021vyx, Drukker:2022beq}.

The M2-branes in $AdS_7\times S^4$ are naturally also homogeneous along the $x_6$ direction 
and are calibrated by the 3-form
\beq
\phi^\bR =
- \diff x_6\wedge  \omega^\bR\,,
\qquad
\omega^\bR=\sum_{I=1}^5\left(\diff x_I\wedge dy_I \right)\,.
\label{eqn:typeR3form}
\eeq
This form is clearly exact. In fact we can write it in two inequivalent ways
\beq
\phi^\bR= - \diff(x_6 \omega^\bR)
=\diff\left(\sum_{I=1}^5 y_I\,\diff x_I\wedge\diff x_6\right).
\eeq
The second expression is more relevant for us, and using $y_I=yn_I$ 
we find that the action of the M2-brane is equal to
\beq
S= T_{M2}\int\diff x_6\,\diff u\sum_{I=1}^5 y\,n_I\,\partial_u x_I\Big|_{y\to\infty}\,.
\eeq

Plugging in the large $y$ expansion \eqref{large-y} and the asymptotic expression for 
$n_I$ in \eqref{eqn:typeR}, we find
\beq
\label{R-bndryS}
S= \frac{N}{4\pi L^3}
\int\diff x_6\,\diff u
\left(y\,n_I^{(0)} \partial_u x_I^{(0)}
+ L^3 n_I^{(0)} \,\partial_u x_I^{(1)}
+ L^3 n_I^{(1)}\,\partial_u x_I^{(0)}\right)
+\cO(y^{-1})\,.
\eeq
The term proportional to $y$ is the usual divergent term, proportional to the area of the 
surface, which is removed by an appropriate counterterm.

The last term in \eqref{R-bndryS} is proportional to $n_I^{(0)} n_I^{(1)}$ and as mentioned above 
vanishes. We are left with the $n_I^{(0)}\,\partial_u x_I^{(1)}$ terms which using 
\eqref{large-y}, \eqref{eqn:typeR} and integration by parts is
\beq
S_\text{ren}= \frac{N}{4\pi}
\int\diff x_6\,\diff u\,
\frac{\partial_u{x}_I^{(0)}}{|\partial_u x^{(0)}|}\,\partial_u H^I
+\cO(y^{-1})
=-\frac{N}{4\pi}
\int\diff x_6\,\diff u\,|\partial_u x^{(0)}|H^I H^I
+\cO(y^{-1})\,.
\label{R-S-ren}
\eeq

Note that for a curve of Type-$\bR$ the mean curvature vector is simply the 
curvature of the curve $\partial_u^2x_\mu^{(0)}$ (in the unit speed parametrisation).

For example, a circle of radius $R$ is given in the unit speed parametrisation 
by $x^{(0)}(u)=(R\cos(u/R),R\sin(u/R))$ and $H^2 = R^{-2}$. The action is then
\beq
  S_\text{ren}=
  -\frac{N}{4\pi} \int_{-D/2}^{D/2} \diff x^6\int_0^{2\pi R} \diff u\, \frac{1}{R^2}
  =-\frac{N}{2}\frac{D}{R}\,,
\eeq
where $D$ is a cutoff on the $x_6$ coordinate. 
This agrees precisely with the result of~\cite{Drukker:2021vyx} derived there 
by solving the minimal surface equations of motion.

If we dimensionally reduce the theory along $x_6$ we find a line operator in 5d along 
the curve $x_I$. If that curve is at a fixed value of $x_5$, we can further reduce to 
4d where we find a BPS Wilson loop in $\cN=4$ SYM. These Wilson loops with the exact 
relation~\eqref{eqn:typeR} were found by Zarembo in~\cite{zarembo:2002an}. 
They are known to have vanishing expectation 
value~\cite{Guralnik:2003di, Guralnik:2004yc, Dymarsky:2006ve}. This is consistent 
with the calculation above, since in this limit $D$, the length of the $x_6$ 
direction vanishes. Thus the 6d observables have non-trivial expectation value 
per unit length, while the 4d ones are trivial.

\subsection{Type-$\bH$}
\label{sec:H}

Type-$\bH$ surfaces are formed of any oriented surface in $\bR^4$ accompanied by the 
scalar coupling~\cite{Drukker:2020bes}
\beq
n_I^{(0)} = \frac{1}{2} \eta_{I \mu\nu} \partial_a x_\mu^{(0)} \partial_b
x_\nu^{(0)} \varepsilon^{ab}
=\frac{1}{2}\epsilon_{IJK} \partial_a x_J^{(0)} \partial_b x_K^{(0)} \varepsilon^{ab}
+\partial_a x_I^{(0)} \partial_b x_4^{(0)} \varepsilon^{ab}
\,,
\label{eqn:typeH}
\eeq
where $\eta_{I \mu\nu}$ is the 't~Hooft chiral symbol expressed on the right 
in terms of the 3d epsilon symbol (with $I,J,K\in\{1,2,3\}$) and
$\varepsilon^{ab}$ is the 2d epsilon tensor (including a factor $h^{-1/2}$). 
This has a nice interpretation in terms of the Gauss map from the surface to its 
tangent space in $S^2\times S^2$. The vector $n_I^{(0)}$ is a coordinate on one of those 
spheres, so this is the projection of the Gauss map to a single $S^2$.

The case when the surface is a cone over a curve on $S^3$, or more precisely 
its conformal transform to a surface in $\bR\times S^5$ has 
been described and studied in great detail in \cite{mezei:2018url}.

The holographic dual of surface operators of Type-$\bH$ are M2-branes restricted 
to an $AdS_5\times S^2$ subspace and calibrated with respect to the form
\bal
\label{eqn:typeH3form}
\phi^\bH &=
\frac{1}{2}\eta_{I \mu\nu}\,\diff x_\mu \wedge \diff x_\nu \wedge dy_I
- \left( \frac{L}{y} \right)^3 \diff y_1\wedge\diff y_2\wedge\diff y_3\,.
\eal
Using spherical coordinates $y$, $\vartheta$ and $\varphi$ instead of $y_I$, we can see that 
this form is exact
\beq
\phi^\bH =
\diff \left[\frac{1}{2} y_I\eta_{I \mu\nu}\,\diff x_\mu \wedge \diff x_\nu 
- L^3\ln y\,\diff \Omega^2\right],
\qquad
\diff\Omega^2=\sin\vartheta\,\diff\vartheta\wedge\diff\varphi\,.
\eeq

The classical action is again a total derivative
\beq
\label{H-bndryS}
\begin{aligned}
  S &=
  T_{M2}\int_V \phi^\bH\\
  &=
  T_{M2} y \int_{\Sigma} \frac{1}{2} n_I^{(0)}
  \eta_{I \mu\nu} \partial_a x_\mu^{(0)} \partial_b x_\nu^{(0)}
  \varepsilon^{ab} \sqrt{h}\, \diff^2 \sigma
  -T_{M2}L^3\ln{y}\int_{\Sigma} \diff\Omega+O(y^0)\,.
\end{aligned}
\eeq
The first term is the usual area divergence that can be cancelled by a counter-term. The 
second one is logarithmically divergent, signifying a conformal anomaly. The integral is over the 
2-sphere at $y_I\to\infty$ and is the same as that appearing in the Gauss map. The integral is 
equal to $4\pi$ times the degree $\nu$ of the map, giving a total anomaly 
\beq
S = 8\pi T_{M2}L^3\nu \log\epsilon + \mathrm{finite} = 2N\nu \log\epsilon +
\mathrm{finite}\,.
\eeq
Here we identified $y \sim \epsilon^{-2}$, where $\epsilon$ a short distance cutoff in the 
CFT. 
This expression was already found in~\cite{Drukker:2020bes} relying on the original 
surface anomaly calculation of Graham and Witten~\cite{graham:1999pm}.

In cases without an anomaly we can evaluate the finite term by again looking at 
$1/y$ corrections to the first term in \eqref{H-bndryS}.
Using \eqref{x_1}, We find
\beq
S_\text{ren} =
\frac{N}{4\pi}\int_{\Sigma} \sqrt{h}\, \varepsilon^{ab}\eta_{I \mu\nu} 
\left(n_I^{(0)} \partial_a x_\mu^{(0)} \partial_b H_\nu
+\frac{1}{2}n_I^{(1)} \partial_a x_\mu^{(0)} \partial_b x_\nu^{(0)}
\right) \diff^2 \sigma
+\cO(y^{-1})\,.
\eeq
The second term is proportional to $n_I^{(1)} n_I^{(0)}$ and therefore vanishes. Plugging the 
expression for $n_I^{(0)}$ \eqref{eqn:typeH}, using the orthogonality relation for 
$\eta_{I \mu\nu}$ and integration by parts we find
\bal
S_\text{ren}
&=
\frac{N}{4\pi}\int_{\Sigma} \sqrt{h}\,\varepsilon^{ab}\varepsilon^{cd}
\eta_{I \mu\nu} \eta_{I \rho\sigma} 
\partial_c x_\rho^{(0)} \partial_d x_\sigma^{(0)}
\partial_a x_\mu^{(0)} \partial_b H_\nu
\diff^2 \sigma
+\cO(y^{-1})
\\&=
\frac{N}{4\pi}\int_{\Sigma} \sqrt{h}\,h^{ab}
\partial_a x_\nu^{(0)} \partial_b H_\nu
\diff^2 \sigma
+\cO(y^{-1})
=-\frac{N}{4\pi}\int_{\Sigma} \sqrt{h}
H_\nu H_\nu
\diff^2 \sigma
+\cO(y^{-1})\,.
\eal
Interestingly, this is exactly the same expression as in the case of Type-$\bR$ surfaces 
\eqref{R-S-ren}.

In cases with an anomaly, the expectation value depends on the choice of
$\epsilon$ so is not well-defined. However since the anomaly is topological, the
ratio of expectation values of two surface operators with the same topology is a
well-defined quantity. It is given by the difference in action, which as above
is simply the difference in integrated mean curvature.

\subsection{Type-$\bC$}
\label{sec:C}

Surfaces of Type-$\bC$ are holomorphic with respect to the usual complex structure 
on $\bR^6=\bC^3$. They have a constant $n_I^{(0)}=\delta_{I1}$. Their holographic duals are 
calibrated with respect to~\cite{Drukker:2020bes}
\beq
\phi^\bC
=(\diff x_1\wedge\diff x_2+\diff x_3\wedge\diff x_4+\diff x_5\wedge\diff x_6)\wedge \diff y_1\,.
\eeq
Clearly this integrates to
\beq
\Phi^\bC
=y_1(\diff x_1\wedge\diff x_2+\diff x_3\wedge\diff x_4+\diff x_5\wedge\diff x_6)\,.
\eeq
In this case holomorphicity ensures that $x_\mu^{(1)}=H_\mu=0$, 
so the action written in terms of complex 2d and 6d coordinates is
\beq
S=T_{M2}\,y\int_\Sigma\,\diff\sigma\,\diff\bar\sigma\,\partial_\sigma z_i\partial_{\bar\sigma}\bar z_i
\eeq
This is simply the area divergence, so the renormalized action vanishes. Note that at 
subleading orders in $N$, surfaces of Type-$\bC$ may have an anomaly~\cite{Drukker:2020bes}.

\section{Lagrangian surfaces}

We shift now to focus on a subclass of Type-$\bH$ surfaces, referred to as Type-L which 
are Lagrangian in $\bR^4$ with the symplectic form 
$\omega_0=\diff x_1\wedge\diff x_2+\diff x_3\wedge\diff x_4$, meaning that
\beq
\omega_0|_{\Sigma} = 0\,.
\eeq
One can see that in this case 
$n_3^{(0)}=0$ \eqref{eqn:typeH}, so the image of the projection of the Gauss map is a circle 
inside $S^2$.

Such surfaces preserve two superchages, which is double the number of a generic Type-$\bH$ 
surface. An easy way to see this, is that these surfaces are also BPS under an alternative 
Type-$\overline\bH$ construction with $n_3^{(0)}\to-n_3^{(0)}$. As such, the M2-branes describing them 
in $AdS_7\times S^4$ are calibrated with respect to two different forms, 
$\phi^\bH$ in \eqref{eqn:typeH3form} and $\phi^{\overline\bH}$ with $\diff y_3\to-\diff y_3$. 
Being calibrated with respect to both, we can define 
\beq
\phi^\bL= \frac{1}{2} \left( \phi^\bH + \phi^{\overline\bH} \right),
\eeq
and then
\begin{align}
\label{phiLpull}
\phi^\bL|_V=\vol_V\,, 
\qquad
\left(\phi^{\bH} - \phi^{\overline\bH} \right)|_V = 0\,.
\end{align}
From \eqref{eqn:typeH3form} we can get the explicit expressions
\begin{align}
\label{phiL}
  \phi^\bL &=
  (\diff x_1 \wedge \diff x_4 + \diff x_2 \wedge \diff x_3) \wedge \diff y_1
  + (\diff x_3 \wedge \diff x_1 + \diff x_2 \wedge \diff x_4) \wedge \diff
  y_2\,,\\
  \phi^\bH - \phi^{\overline\bH} &=
  2\left( \diff x_1 \wedge \diff x_2 + \diff x_3 \wedge \diff x_4 -
  \left(\frac{L}{y}\right)^3 \diff y_1 \wedge \diff y_2 \right) \wedge \diff y_3\,.
  \label{eqn:symplecticformphi}
\end{align}
Since the pullback of $\phi^\bL$ is the volume form, we conclude that the
M2-brane is in the subspace of $AdS_7 \times S^4$ at constant $x_5$, $x_6$, $y_{3,4,5}$.

This is an even dimensional manifold and it is natural to associate a complex
structure. Taking the usual complex structure $J$ for $z_1 = x_1 + i x_2$, $z_2 = x_3
+ i x_4$, $z_3 = -(y_2 + i y_1)$ and the symplectic form $\omega$ to be the
extension of $\omega_0$ appearing in~\eqref{eqn:symplecticformphi}
\bal
\label{omega-lag}
\omega &= \diff x_1 \wedge \diff x_2 + \diff x_3 \wedge \diff x_4 -
\left( \frac{L}{y} \right)^3 \diff y_1 \wedge \diff y_2\\
&= \frac{i}{2} \left[ \diff z_1 \wedge \diff \bar{z}_1 + \diff z_2 \wedge \diff \bar{z}_2
+ \left( \frac{L}{y} \right)^3 \diff z_3 \wedge \diff \bar{z}_3 \right],
\eal
this defines a K\"ahler manifold with metric
\begin{align}
\tilde{G} 
= \diff x_\mu \diff x_\mu + \left( \frac{L}{y} \right)^3 \diff y_I\diff y_I
= \diff z_1 \diff \bar z_1+\diff z_2 \diff \bar z_2 
+ \left( \frac{L}{y} \right)^3 \diff z_3\diff \bar z_3\,.
\end{align}
The metric $\tilde{G}$ is not the same as $G$ in \eqref{eqn:metric}. It is related to it 
(on an appropriate subspace) by the conformal transformation
\begin{align}
  G = \frac{y}{L} \tilde{G}\,.
\end{align}
In addition, an (almost) K\"ahler manifold has a holomorphic 3-form
\begin{align}
  \Omega = \diff z_1 \wedge \diff z_2 \wedge \diff z_3\,,
\end{align}
related to $\omega$ as
\begin{align}
  \frac{1}{3!}\left( \frac{y}{L} \right)^3 \omega^3
  =
  -\left( \frac{i}{2} \right)^3 \Omega \wedge \bar{\Omega}
  =
  -\diff x_1 \wedge \diff x_2 \wedge \diff x_3 \wedge \diff x_4 \wedge \diff y_1
  \wedge \diff y_2\,.
\end{align}
This makes this space an almost Calabi-Yau/special K\"ahler manifold
according to~\cite{joycespecial}.

We now prove that the M2-brane world-volume $V$ is a special Lagrangian submanifold, which 
requires
\begin{align}
  \Im{\Omega}|_V = 0\,, \qquad
  \omega|_V = 0\,.
  \label{eqn:speciallagrangian}
\end{align}
By expanding $\Omega$ we find
\begin{align}
\label{Omega-lag}
  \Omega = 
  \frac{1}{2}\sum_{i,j=1}^2\eta_{i \mu\nu} (\delta_{ij} - i \varepsilon_{ij})\,
  \diff x_\mu \wedge \diff x_\nu \wedge \diff y_j\,.
\end{align}
So its real part is $\Re\Omega=\phi^\bL$ \eqref{phiL}. Since the M2-brane are
calibrated by $\phi^\bL$, the pullback of $\Re{\Omega}$ is the volume form, so the
M2-branes are parallel to it. From \eqref{Omega-lag} it is clear that 
$\Im{\Omega}$ is orthogonal to
$\Re{\Omega}$, so the pullback $\Im{\Omega}|_V= 0$ vanishes.
Finally note that $\omega = \partial_{y_3} \cdot\phi^\bH=\partial_{y_3} \cdot\phi^{\overline\bH}$, 
so $\omega|_V=0$ by virtue of the second equation in \eqref{phiLpull}. 
This proves 
\eqref{eqn:speciallagrangian} and therefore $V$ is a special Lagrangian submanifold.

\subsection{Lagrangian surfaces with $U(1)$ symmetry}

If the surface has a symmetry under $(z_1,z_2)\to (e^{i\chi} z_1,e^{-i\chi}z_2)$, 
we expect the M2 world volume to have the same symmetry, which then matches 
the conditions on Lagrangian surfaces studied 
in~\cite{joycesymmetries, joyceU1}.

Following~\cite{joyceU1}, we define $w=\Im(z_1z_2)=x_1x_4+x_2x_3$ and 
$v=\Re(z_1z_2)=x_1x_3-x_2x_4$, then based on 
the symmetry we can parametrise the surface in terms of $y_2$, $w$ and an angular 
direction such that the nontrivial dependence is
\beq
y_1(y_2,w)\,,
\qquad
v(y_2,w)\,.
\eeq
The conserved quantity associated with the $U(1)$ symmetry is
\beq
\label{u1-constraint}
|z_1|^2-|z_2|^2=2a\,.
\eeq
with a constant $a$.

In these coordinates it's easy to check that $\omega$ \eqref{omega-lag} takes the form
\begin{align}
  \omega &=
  \frac{\diff v \wedge \diff w}{|z_1|^2 + |z_2|^2} -
  \frac{L^3}{y^3} \diff y_1^2 \wedge \diff y_2^2\,.
\end{align}
In this form the constraint $\omega|_V = 0$ is straightforward to read, and
with a bit more work one can obtain the constraint $\Im\Omega|_V=0$ as well.
They respectively give 
\beq
\frac{\partial v}{\partial y_2}
=-2\sqrt{w^2+v^2+a^2}\frac{L^3}{y^3}\frac{\partial y_1}{\partial w}\,,
\qquad
\frac{\partial v}{\partial w}
=\frac{\partial y_1}{\partial y_2}\,.
\label{CR}
\eeq
Note that $2\sqrt{w^2+v^2+a^2}=|z_1|^2+|z_2|^2$. These can be thought of as modified 
Cauchy-Riemann equations and are related to the equations in~\cite{joyceU1} by 
the factor of $L^3/y^3$.

Alternatively, the $U(1)$ symmetry can act on $(z_1,z_3)\to (e^{i\chi} z_1,e^{-i\chi}z_3)$. 
Adapting the case above, we define $w'=-\Im(z_1z_3)=x_1y_1+x_2y_2$ and 
$v'=-\Re(z_1z_3)=x_1y_2-x_2y_1$, then we parametrise the surface in terms of $x_3$, $w$ and an angular 
direction such that the nontrivial dependence is
\beq
x_4(x_3,w')\,,
\qquad
v'(x_3,w')\,.
\eeq
Now the analog of the constraint \eqref{u1-constraint} is
\beq
\label{constraint2}
|z_1|^2+\frac{2L^3}{\sqrt{|z_3|^2+y_3^2}}=|z_1|^2+\frac{2L^3}{y}=2a\,,
\eeq
and $\omega|_V=\Im\Omega|_V=0$ imply
\beq
\frac{\partial v'}{\partial x_3}
= -\left(2 a \frac{y^3}{L^3}-y^2 -y_3^2 \right)
\frac{\partial x_4}{\partial w'}\,,
\qquad
\frac{\partial v'}{\partial w'}
=\frac{\partial x_4}{\partial x_3}\,.
\label{CR2}
\eeq
The parenthesis in the first equation is simply $y^3|z_1|^2/L^3 +|z_3|^2$.
It can be expressed in terms of $v$, $w$, $a$ only
by recasting the expression for $a$ as a cubic equation for $y$
\beq
2a = \frac{v^2+w^2}{y^2-y_3^2} + \frac{2L^3}{y}\,.
\eeq

\subsubsection{Simplest solutions}
\label{sec:simplest}
The simplest way to solve \eqref{CR} is if the left equation vanishes identically, 
so
\beq
v=\alpha w+\beta\,,
\qquad
y_1=\alpha y_2+\gamma\,.
\eeq
In this family of solutions, the coordinates $z_1$ and $z_2$ do not depend on the 
$y_i$ coordinates, as is also the case 
for holomorphic surfaces in Section~\ref{sec:C}. This is not 
a coincidence, as for large $y$, the surface approaches a single point with 
$y_1/y_2=\alpha$, exactly as in the case of surfaces of Type-$\bC$. 
To see the relation, define the complex coordinates
\beq
\zeta=x_1+\frac{i}{\sqrt{1+\alpha^2}}(x_3-\alpha x_4)\,,
\qquad
\eta=x_2-\frac{i}{\sqrt{1+\alpha^2}}(x_4+\alpha x_3)\,.
\eeq
Then the equations
\beq
x_1^2+x_2^2-x_3^2-x_4^2=2a\,,
\qquad
x_1x_3-x_2x_4=\alpha(x_1x_4+x_2x_3)+\beta\,,
\eeq
combine into the holomorphic quadratic equation 
$\zeta^2+\eta^2=2a+2i\beta/\sqrt{1+\alpha^2}$.

This solution is more general than that in Section~\ref{sec:C} because of the 
constant $\gamma$. When it is different from zero, the surface does not reach
$y_1=y_2=0$ and instead continues to negative values of $y_1$ and $y_2$ and another 
asymptotic region, representing the correlation function of two coincident surfaces 
with opposite orientation and scalar couplings. This is a rather singular configuration, 
where the same boundary conditions have degenerate solutions with arbitrary $\gamma$.
These solutions are similar in spirit to those studied in~\cite{Klebanov:2006jj}, and are also related to zig-zag symmetry, see for 
example~\cite{Polyakov:1997tj,drukker:1999zq,Polyakov:2000ti}.
We find more such degeneracies below in Section~\ref{sec:torus}.

The same can be done with \eqref{CR2}, where now
\beq
\label{lin-w'x4}
v'=\alpha w'+\beta\,,
\qquad
x_4=\alpha x_3+\gamma\,.
\eeq
This simple dependence of $x_3$ and $x_4$ is a version of Type-$\bR$, see 
section~\ref{sec:R}.

The remaining coordinates satisfy
\beq
x_1y_2-x_2y_1=\alpha(x_1y_1+x_2y_2)-\beta\,,
\qquad
x_1^2+x_2^2+\frac{1}{2y}=2a\,.
\eeq
For $\beta=0$, this is up to rotations the cylinder solution of~\cite{Drukker:2021vyx} reviewed in 
Appendix~\ref{app:cylinder}.
The case $\beta \neq 0$ has a similar interpretation as the degenerate solutions 
for two coincident surfaces discussed above.

\subsubsection{Torus}
\label{sec:torus}

Example 5.1 in~\cite{joyceU1}, which was first presented in Section III.3.A 
of~\cite{harvey1982calibrated}, has $U(1)^2$ symmetry. It is the intersection of 
the two construction above, so
\beq
|z_1|^2-|z_2|^2=2a\,,
\qquad
|z_1|^2+\frac{2L^3}{y}=2b\,,
\qquad
\Im(z_1z_2z_3)=0\,.
\eeq
The third condition (up to a constant) is a consequence of the second of the equations 
in \eqref{CR} and \eqref{CR2}, which can be seen by writing $z_1$, $z_2$ and 
$z_3$ in polar coordinates. 
From the limit $y\to\infty$ we find that both $|z_1|$ and $|z_2|$ are constants, 
making a torus with radii squared
$R_1^2=2b$ and $R_2^2=R_1^2-2a$. The third equation gives the relation between 
the three phases. This completely specifies the surface and one can verify that 
the differential equations above are indeed satisfied.

One still needs to check whether the solution is regular. An obvious condition is 
$R_1^2>0$, $R_2^2>0$. Without loss of generality we can assume $R_1\leq R_2$. Then 
if we impose $b=L^3/y_3$, then at $z_3=0$ the surface closes off smoothly at 
$z_1=0$. This is the solution for the toroidal surface operator found 
in~\cite{Drukker:2021vyx} and reviewed in Appendix~\ref{app:torus}.

For $b<L^3/y_3$ the surface cannot extend to $z_3=0$ and for $b>L^3/y_3$ it cannot 
reach $|z_1|=0$. In the former case we can extend the surface to cover the 
$|z_1|^2\leq2b$ disc twice, while reaching a minimal value of 
$|z_3|^2=L^6/b^2-y_3^2$ at $z_1=0$. This surface has two asymptotic regions with 
$y\to\infty$, so is the correlation function of two coincident surface operators 
with opposite orientation but opposite scalar couplings that allow them to be mutually BPS. 
For the second case, with $b>L^3/y_3$, the surface is a double cover of the 
$z_3$ plane, but covers only annuli in the $z_1$ and $z_2$ planes (also twice).

In these cases we find the same tori at large $y$ irrespective of the value of $y_3$, 
so the solutions are degenerate. This resembles the situation studied 
in~\cite{Klebanov:2006jj}.

\subsubsection{Generalised cylinder}

Consider the limit of $a\to-\infty$ in \eqref{u1-constraint}. This means that $|z_1|\ll|z_2|$ 
and if we look at the $U(1)$ action for small $\chi$, it is essentially a translation along 
the tangent in the $z_2$ direction and does not act on $z_1$.  Assume $x_4\sim
\sqrt{2|a|}$ and $x_3\sim0$, 
then the surface is locally a curve in the $z_1$ plane times the $x_3$ direction, so a special 
class that is both Lagrangian and of Type-$\bR$ (one can easily verify that the ans\"atze for 
$n_I^{(0)}$ agree).

Now we have $w\sim \sqrt{|a|} x_1$ and $v\sim- \sqrt{|a|} x_2$. Equation \eqref{CR} then becomes
\beq
\label{gen-cyl}
\frac{\partial x_2}{\partial y_2}
=\frac{L^3}{y^3}\frac{\partial y_1}{\partial x_1}\,,
\qquad
\frac{\partial x_2}{\partial x_1}
=-\frac{\partial y_1}{\partial y_2}\,.
\eeq
These equations can also be derived directly from $\Im\Omega|_V=\omega_V=0$ when we impose 
the extra constraint $x_4=0$. It is not too hard to check that they are satisfied by the 
crease solution of~\cite{Drukker:2022beq} reviewed in Appendix~\ref{app:crease}.

One can try to take a similar limit on the other $U(1)$ symmetric configuration 
\eqref{constraint2}, such that there is translation invariance in the $y_I$ directions. 
While the limit is a bit more subtle, the equations \eqref{CR2} should be the regular 
Cauchy-Riemann equations in $\bR^4$. This reproduces the results from Section~\ref{sec:C}, 
where the surfaces were homogeneous in the $y_I$ directions.

\subsection{Perturbed cylinder}
\label{sec:pert-cyl}

The cylinder solution in Appendix~\ref{app:cylinder} solves the generalised 
Cauchy-Riemann equation \eqref{gen-cyl}. Following Joyce~\cite{JOYCE200535}, 
we can study solutions close to it 
by taking a function $f(y_2, x_1)$ and taking
\beq
y_1\to y_1-\frac{\partial f}{\partial x_1}\,,
\qquad
x_2\to x_2+\frac{\partial f}{\partial y_2}\,.
\eeq
Equation \eqref{gen-cyl} now gives
\beq
\frac{\partial x_2}{\partial y_2}-\frac{L^3}{y^3}\frac{\partial y_1}{\partial x_1}
=\frac{\partial^2 f}{\partial y_2^2}+\frac{L^3}{y^3}\frac{\partial^2 f}{\partial x_1^2}
=0\,.
\eeq
Alternatively, this can be written as
\beq
L^3\frac{\partial^2 f}{\partial x_1^2}
+\left(\left(y_1-\frac{\partial f}{\partial x_1}\right)^2+y_2^2+y_3^2\right)^{3/2}
\frac{\partial^2 f}{\partial y_2^2}
=0\,.
\eeq
Here $y_1$ is as in the cylinder solution \eqref{cylsol}, which is an explicit function 
of $x_1$ and $y_2$. For small $f$ we can linearise the equation to
\beq
L^3\frac{\partial^2 f}{\partial x_1^2}
+\left(y_1(x_1,y_2)^2+y_2^2+y_3^2\right)^{3/2}
\frac{\partial^2 f}{\partial y_2^2}
=0\,.
\eeq
In this equation $y_1$ is the original solution we are perturbing about, so 
generally it is not a separable equation. It would still be nice to find examples 
where it can be solved explicitly. Other results on special Lagrangian 
manifolds could also lead to new M2-brane solutions.

\section*{Acknowledgements}
We are grateful to J. Maldacena, M. Martone, D. Panov and especially  M. Probst
for helpful discussions.  ND would like to thank l'\'Ecole Polytechnique
F\'ed\'erale de Lausanne, the Simons Center for Geometry and Physics, Stony
Brook and the KITP, Santa Barbara for their hospitality in the course of this
work. ND's research is supported by the Science Technology \& Facilities council
under the grants ST/T000759/1 and ST/P000258/1 and by the National Science
Foundation under Grant No. NSF PHY-1748958.  MT gratefully acknowledges the
support from the Institute for Theoretical and Mathematical Physics (ITMP,
Moscow) where this project began, and Université Laval, the Simons Center for
Geometry and Physics, Stony Brook and New York University where parts of this
project were realised.  MT's research is funded by the Engineering \& Physical
Sciences Research Council under the grant EP/W522429/1.  Research at Perimeter
Institute is supported by the Government of Canada through the Department of
Innovation, Science and Economic Development and by the Province of Ontario
through the Ministry of Research and Innovation.

\appendix
\section{Review of known BPS solutions}
\label{app:solutions}
\subsection{Torus}
\label{app:torus}
The torus solution was presented in~\cite{Drukker:2021vyx}. The world volume coordinates are 
$\rho$, $\varphi_1$ and $\varphi_2$ and in the metric \eqref{eqn:metric} it takes the form
\bal
\label{torusansatz}
x_1&=r_1(\rho)\cos\varphi_1\,,
\qquad
&
x_2&=r_1(\rho)\sin\varphi_1\,,
\\
x_3&=r_2(\rho)\cos\varphi_2\,,
\qquad
&
x_4&=r_2(\rho)\sin\varphi_2\,,
\\
y_1&=-\rho\sin(\varphi_1+\varphi_2)\,,
\qquad
&
y_2&=\rho\cos(\varphi_1+\varphi_2)\,,
\eal
with
\beq
\label{torussol}
r_1^2(\rho)=R_1^2-\frac{2L^3}{y}\,,
\qquad
r_1^2(\rho)=R_2^2-\frac{2L^3}{y}\,,
\qquad
y_3=\frac{2L^3}{R_1^2}\,.
\eeq
Here $y^2=y_1^2+y_2^2+y_3^2=\rho^2+y_3^2$ and $R_1$ and $R_2$ are constants, the radii of 
the torus at the boundary $y\to\infty$ and we assume $R_1\leq R_2$.

\subsection{Cylinder}
\label{app:cylinder}

The cylinder is a limit of the torus, where effectively $R_2\to\infty$. We take 
$\rho$, $\varphi_1$ and $v$ as worldvolume coordinates and
\bal
\label{cylansatz}
x_1&=r_1(\rho)\cos\varphi_1\,,
\qquad
&
x_2&=r_1(\rho)\sin\varphi_1\,,
\\
x_3&=0\,,
\qquad
&
x_4&=v\,,
\\
y_1&=-\rho\sin\varphi_1\,,
\qquad
&
y_2&=\rho\cos\varphi_1\,,
\eal
with
\beq
\label{cylsol}
r_1^2(\rho)=R_1^2-\frac{2L^3}{y}\,,
\qquad
y_3=\frac{2L^3}{R_1^2}\,.
\eeq

\subsection{Crease}
\label{app:crease}

The crease solution was found in~\cite{Drukker:2022beq}. In the BPS case, if we take 
$r$, $u$ and $v$ as worldvolume coordinates, it can be expressed as
\beq
\begin{gathered}
x_1=r\cos\varphi(u)\,,
\qquad
x_2=r\sin\varphi(u)\,,
\qquad
x_3=\sqrt{1+u^2}v\,,
\qquad
x_4=0\,,
\\
y_1=\frac{4L^3}{u^2r^2}\sin\vartheta(u)\,,
\qquad
y_2=-\frac{4L^3}{u^2r^2}\cos\vartheta(u)\,,
\qquad
y_3=0\,,
\end{gathered}
\eeq
with
\beq
\tan\vartheta=\frac{2J\sqrt{1+u^2-J^2u^4}}{1+(1+2J^2)u^2}\,,
\qquad
\sin(\varphi-\vartheta)=\frac{Ju^2}{\sqrt{1+u^2}}\,.
\eeq
$J$ is an arbitrary constant, leading to a one-parameter family of solutions.

It is simple to check that this also implies
\beq
\frac{\sin\vartheta}{\cos\varphi}=\frac{2J}{\sqrt{1+u^2}}\,,
\eeq
where in our notation $u=2L^{3/2}/r\sqrt{y}$. 
This can also be written as
\beq
r\cos\varphi-\frac{1}{J}\sqrt{\frac{r^2}{4}+\frac{L^3}{y}}\sin\vartheta=0\,.
\eeq

\bibliographystyle{utphys2}
\bibliography{ref}

\end{document}